\documentclass[conference]{IEEEtran}
\IEEEoverridecommandlockouts
\usepackage{cite}
\usepackage{amsmath,amssymb,amsfonts}
\usepackage{algorithmic}
\usepackage{graphicx}
\usepackage{textcomp}
\usepackage{xcolor}
\usepackage{subcaption}
\usepackage{booktabs}
\usepackage{url}
\usepackage{amsmath}

\usepackage{graphicx}
\usepackage{caption}
\usepackage{subcaption}
\def\BibTeX{{\rm B\kern-.05em{\sc i\kern-.025em b}\kern-.08em
    T\kern-.1667em\lower.7ex\hbox{E}\kern-.125emX}}
\begin{document}

\title{Unveiling User Engagement Patterns on Stack Exchange Through Network Analysis}

\author{
    \IEEEauthorblockN{1\textsuperscript{st} Agnik Saha}
    \IEEEauthorblockA{
        \textit{Department of Computer Science} \\
        \textit{Georgia State University}\\
        Atlanta, Georgia, USA \\
        asaha8@gsu.edu}
    \and
    \IEEEauthorblockN{2\textsuperscript{nd} Mohammad Shahidul Kader}
    \IEEEauthorblockA{
        \textit{Individual Researcher} \\
        Atlanta, Georgia, USA \\
        mzk0089@auburn.edu}
    \and
    \IEEEauthorblockN{3\textsuperscript{rd} Mohammad Masum}
    \IEEEauthorblockA{
        \textit{Department of Applied Data Science} \\
        \textit{San Jose State University}\\
        San Jose, USA \\
        mohammad.masum@sjsu.edu}
}

\maketitle

\begin{abstract}
Stack Exchange, a question-and-answer(Q\&A) platform, has exhibited signs  of a declining user engagement. This paper investigates user engagement dynamics across various Stack Exchange communities including Data science, AI, software engineering, project management, and GenAI. We propose a network graph representing users as nodes and their interactions as edges. We explore engagement patterns through key network metrics including Degree Centerality, Betweenness Centrality, and PageRank. The study findings reveal distinct community dynamics across these platforms, with smaller communities demonstrating more concentrated user influence, while larger platforms showcase more distributed engagement. Besides, the results showed insights into user roles, influence, and potential strategies for enhancing engagement. This research contributes to understanding of online community behavior and provides a framework  for future studies to improve the Stack Exchange user experience. 

\end{abstract}


\section{Introduction}
Question-and-answer(Q\&A) platforms have emerged as essential tools for online knowledge exchange and problem-solving because they allow users to ask questions, respond to inquiries, and have thoughtful conversations \cite{andoh2024integrated}. A number of Q\&A websites such as Quora, Stack Exchange network, and Reddit's Ask Me Anything (AMA) threads are helpful tools for people searching for information online on diverse topics \cite{hadi2022users}. Stack Exchange is a leading platform for Q\&A exchanges while Stack Overflow is the flagship site, dedicated to computer programmers \cite{ndukwe2022perceptions}. Q\&A platforms are widely recognized for their usefulness and popularity, however, a recent decline in user engagement is prominent \cite{mustafa2024share}. For example, Stack Overflow saw an 11\% decline in non-deleted questions and a 12.9\% decline in new user registrations in Jan 2023 \cite{stackexchange_decline}. 

This potential decline in Q\&A platforms such as Stack Exchange is intriguing for several reasons. First, the advent of AI tools such as ChatGPT has provided users with an alternative means of obtaining answers to their technical and non-technical questions \cite{kabir2024stack}. For example, ChatGPT answers are preferred by programmers 35\% of the time despite being incorrect 52\% of the time due to their comprehensiveness and language models \cite{kabir2024stack}. The increasing reliance on AI-driven bots, and the homogenization of knowledge might stifle creativity; ultimately limit the range of options accessible to technical professionals \cite{caporusso2023generative}.  Second, sometimes, stringent moderation guidelines and severe community reactions impede the expansion of new users. For instance, the decline in user engagement on Stack Overflow means fewer opportunities for novice programmers to develop their problem-solving skills and critical thinking abilities \cite{caporusso2023generative}.  Besides, higher likelihood of questions being duplicated can discourage users from posting new questions or from engaging with the platform \cite{luo2024continuous}. Third, the shift of community dynamics including the decline of experienced users (e.g., contributing high-quality answers), the emergence of social media-based specialized communities (e.g., discord community), misalignment with developer workflow (e.g., not integrated with IDEs) leads to a perceived decline in Q\&A platforms \cite{saxena2022users}. Moreover, platforms like Stack Overflow foster a culture of collaborative problem-solving; a decline in engagement diminishes this collaborative spirit, leading to a more transactional and less interactive knowledge-sharing environment \cite{zolduoarrati2024harmonising}. 

Understanding the dynamics of user engagement in stack exchange platforms is essential as contemporary literature has provided little knowledge on the core mechanisms driving  user engagement. Traditional methods for evaluating user engagement often rely on metrics like question views, answer counts, or reputation scores which are quite insufficient in building optimized platform \cite{hadi2022users}. The proposed network analysis approach goes beyond these surface-level indicators and delves deeper into the underlying dynamics of user interaction patterns. 

\textbf{Research questions:} The key research questions that we address as part of this study. \\
\textbf{RQ1:} What are engagement patterns between questioners and responders on Q\&A platforms (i.e., Stack Exchange platform)? \\
\textbf{RQ2}: How can we identify influential users based on interaction frequency, response timeliness, and their interactions with other users on Stack Exchange using a graph mining approach?\\
\textbf{RQ3}: How do the structure and dynamics of various Stack Exchange platforms differ, as indicated by the correlations between centrality measures?  

The current study investigates user engagement patterns and community dynamics on Stack Exchange sites including Data Science, Artificial Intelligence (AI), Generative AI (GenAI), Software Engineering, Project Management,  through network analysis and proposes a novel approach that involves constructing a network graph where users are represented as nodes and their interactions (asking/answering questions) are depicted as edges. By analyzing various network measures (centrality, edge weight), we can gain insights into user behavior and engagement levels. 

To address the above-stated research problems, we constructed and analyzed a (Questioner Responder) QR Network representing interactions within Stack Exchange communities. The network was designed with nodes corresponding to users and edges representing the interactions between them. Edge weights were calculated based on the response time, with faster interactions indicating stronger connections. We then calculated various centrality measures, including Degree Centrality, Betweenness Centrality, Closeness Centrality, PageRank, Eigenvector Centrality, and Harmonic Centrality. These metrics allowed us to identify influential users and examine their roles within the network. Finally, we conducted a comparative analysis of the different Stack Exchange platforms by exploring the correlations between centrality measures and edge weights. This comparison was further supported by statistical summaries including mean and standard deviation of centrality scores, to highlight the structural and dynamic differences across these communities. 

In the following sections, we first review the related work in Section II. Section III outlines our methodology including a description of the dataset and the process of network construction. In Section IV, we present an analysis of the network. Section V describes our findings on the correlation of centrality measures. We discuss the implications of these findings for understanding user engagement and network dynamics in Section VI. Finally, we outline future research directions in Section VII and provide concluding remarks in Section VIII.

\section{Related Work}
\subsection{Activity theory}
In human-computer interaction (HCI) literature, activity theory plays an important role in explaining structure, flow, and attributes of a system that surrounds people and technology \cite{leont1978activity}; \cite{nardi1996context}. According to \cite{kuutti199922}, two facets of activity theory can be described including 1) object-orientedness and 2) hierarchical structure of activity. According to the first facet, user interaction is directly related to subject-object interaction and activities can be differentiated from each other according to their objects \cite{kuutti199922}. In the current context, each user (subject) interacts with specific topics or problems (objects) using stack exchange platforms (tools). Subjects or actors (e.g., questioners, passive users) have specific needs to be fulfilled by an interaction with the surrounding system (e.g., community). The outcome of these interactions can be high-quality answers, upvotes, or designated best answers. Each Q\&A platform has a vast amount of users (community) who might engage in activities aimed at solving problems. As per second facet of hierarchical structure of activity, division of labor (i.e., high active responder, moderate active responder, both ways Q\&R, passive user/watcher, highly active questioner vs low active questioner) determines the transformable activities \cite{kaptelinin2012activity}. The community shares the objective of transforming a posed question into a comprehensive and accurate answer. Rules within the community, both explicit (e.g., platform guidelines) and implicit (e.g., norms of conduct and quality expectations), govern interactions and contributions of members \cite{nardi1996context}. By analyzing the roles of subjects, objects, tools, and community within this framework, we can gain insights into how to enhance participation, improve answer quality, and foster a supportive environment for knowledge sharing.
\subsection{Research on Stack Exchange Platforms}
Several studies have explored different aspects of user engagement and content quality on Stack Exchange platforms. Moutidis et al. \cite{Moutidis:2021} investigated user communities through tag usage trends through trackinng new technologies, user base development, and identifying topic clusters attracting specific user groups. Anderson et al. \cite{Anderson:2012} delved into the dynamics of question answering and voting. They analyzed how reputation influences user participation and the impact of overall question activity on answer quality. Wang et al. (2013) examined the potential segregation between questioners and answerers on Stack Overflow through exploring user communication patterns on the platform. Bachschi et al. \cite{Bachschi:2020} explored user characteristics such as tenure, location, and sub-community involvement to  investigate on how negative feedback affects the transition from question asking to answer provision, ultimately exploring motivations behind user contribution on collaborative platforms. Bhat et al.\cite{Bhat:2014} analyzed the relationship between response time and various factors through tag-based features (e.g., tag frequency) and non-tag features (e.g., code length) to understand how question characteristics influence response speed. Wang et al. (2018) \cite{Wang:2018} employed a logistic regression model to explore factors affecting the time to receive an accepted answer. They analyzed 46 factors across four dimensions (question, questioner, answer, and answerer) on Stack Overflow and other Q\&A platforms. Yazdaninia et al. (2021)\cite{yazdaninia:2021} focused on predicting unanswered questions on Stack Overflow. They utilized tag-based features (e.g., tag popularity) and built a model to identify questions with a low likelihood of receiving an answer.
Mondal et al. (2021)\cite{Mondal:2021} compared the answered and unanswered questions on Stack Overflow. They analyzed both quantitative factors (readability, topic response ratio) and qualitative aspects (problem description, formatting) to predict answerability. Zhang et al. (2021) \cite{Zhang_2:2021} investigated answer obsolescence across all domains on Stack Overflow. They explored how users deal with outdated answers and proposed recommendations for improving answer longevity. Zhang et al. (2022) \cite{Zhang:2021} evaluated the effectiveness of Stack Overflow's comment ranking mechanism. They found it suboptimal for highlighting informative comments due to the prevalence of tie scores. Additionally, they developed a classifier to distinguish between valuable and irrelevant comments. 
Hazra et. al. \cite{hazra2023evaluating} showed us that the response time of the answers in Stack Exchange platform is dependent on metadata, question structure and user interaction patterns. \\
These studies provided valuable insights into user interaction patterns, content quality, and factors influencing user engagement on Stack Exchange platforms. Our work builds upon this existing research by proposing a network analysis approach to delve deeper into user engagement dynamics. 
\subsection{Research on User engagement}
Hasan et. al. \cite{hasan2024measuring} proposed a new method called BehaviourRank Influence (BRI) that focuses on analyzing the behavior of Twitter users instead of the network structure. This research \cite{karamitsos2022graph} proposed a method using graph mining to analyze the strength of connections between users on Twitter for a more accurate measure of user engagement instead of simply looking at the number of interactions a tweet gets. The research \cite{ibrahim2017exploring} aimed to understand how user engagement on Twitter (likes, replies, etc.) affects customer sentiment and brand perception to help businesses improve their social media strategy.  \\
Based on an extensive review of existing literature, our research endeavors to decipher the dynamics of interaction between question seekers and responders, delving deep into the nuances of their engagement patterns. Moreover, our aim extends to the detection and analysis of broader interaction patterns exhibited by users across the platform. Through this study, we aspire to gain insights into the mechanisms through which knowledge flows among users, contributing to a comprehensive understanding of collaborative knowledge sharing within online communities.
\section{Methodology}
\begin{figure*}[h]
  \centering
  \includegraphics[width=0.8\textwidth, height=0.5\textheight]{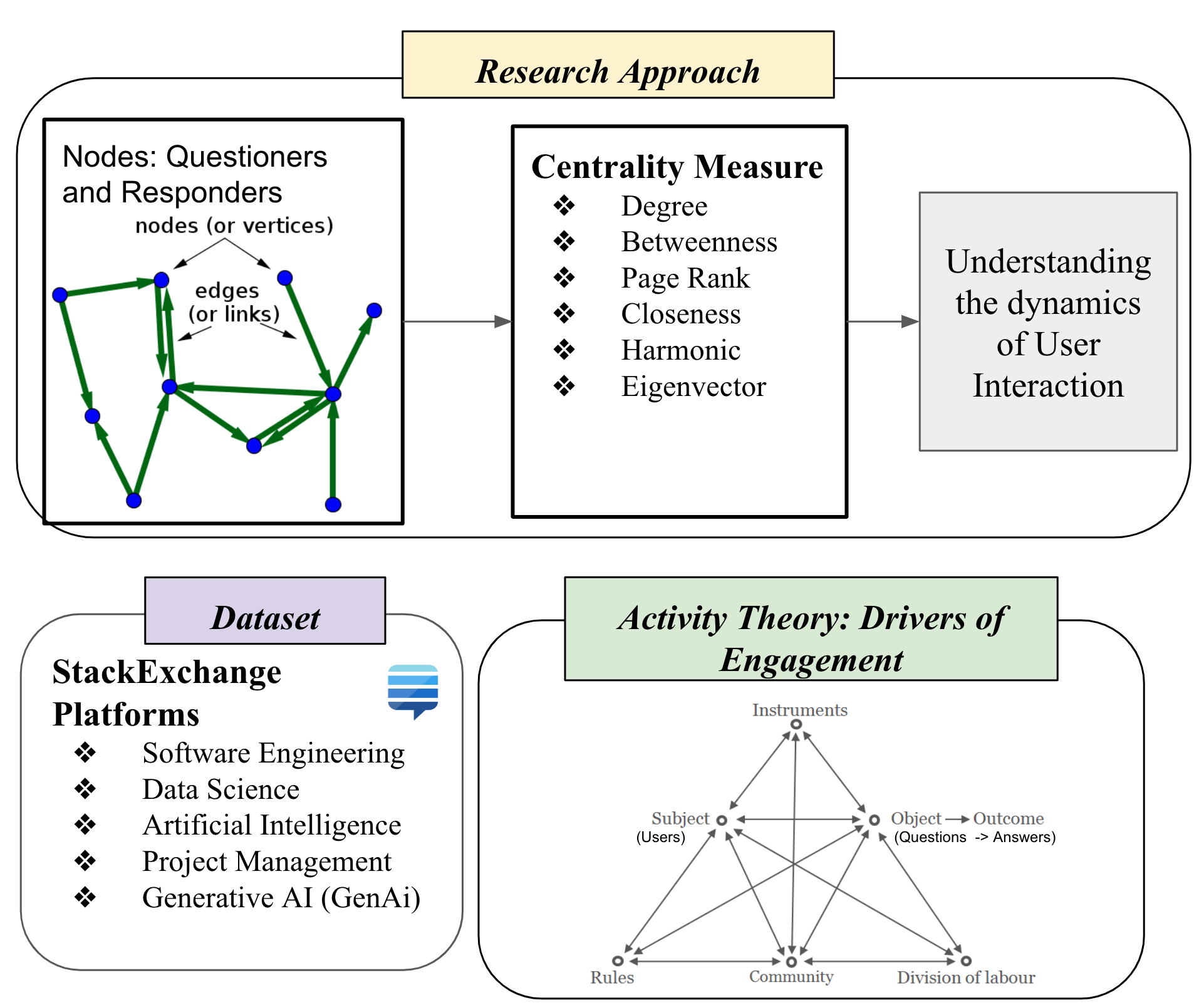}
  \caption{Conceptual Framework for User Engagement on Stack Exchange Platforms, integrating components of Activity Theory with Research Approach and Methodology. This framework illustrates the interactions between users (both questioners and responders), and the overall community structure, emphasizing how these elements influence user engagement and network dynamics.}
  \label{fig:example_image}
\end{figure*}
\subsection{Dataset Preparation and Analysis}
As of March 1st, 2024, there are a total of 175 Q\&A websites within the Stack Exchange family. These websites cover a very wide range of topics, such as technology, culture, art, and business. In this paper, we have conducted experiments on moneyexchange data from the StackExchange data dump. All the datasets contain information till March'2024, starting from the launch date of each individual StackExchange platform. All the data were retrieved from the Archive.org platform \footnote{https://archive.org/download/stackexchange}, which hosts the entire history of every Stack Exchange community, including the tags used to annotate questions and the votes of each question and answer. These archives are anonymized dumps of user-contributed content, formatted as separate XML files compressed using bzip2 compression and updated every three months.
All user content contributed to the Stack Exchange network is cc-by-sa 4.0 \footnote{https://creativecommons.org/licenses/by-sa/4.0/} licensed, intended to be shared and remixed. The gathering, processing, and displaying of these data are all done in a manner that is entirely compliant with the rules and conditions that govern the Stack Exchange network. Posts, Users, Votes, Comments, PostHistory, and PostLinks are all included in each site's archive. \\
Recent studies and analyses have shown a significant decline in user activity across various Stack Exchange platforms, including those focused on Data Science, Artificial Intelligence, Software Engineering, Project Management, and GenAI. This decline has become more pronounced following the advent of advanced AI tools like ChatGPT. Users increasingly turn to AI models like ChatGPT for instant and comprehensive responses, reducing the frequency of posts and engagements on traditional Q\&A platforms. \\
The five Stack Exchange platforms—Data Science, Artificial Intelligence, Software Engineering, Project Management, and GenAI—were carefully selected for this study based on their relevance, community size, and the nature of the discussions they host, which reflect a broad spectrum of technical and managerial topics.

Data Science (launched in July 2014) is a key platform for discussions on algorithms, machine learning, and data analysis, reflecting its critical role in modern technology and hosting over 36,775 questions. Artificial Intelligence (launched in June 2016) deals with cutting-edge topics in AI, making it a vital space for experts and enthusiasts, with over 12,380 questions. Software Engineering (launched in August 2010) is one of the oldest and most comprehensive platforms, addressing software development practices with over 63,423 questions. Project Management (launched in December 2010) offers a unique perspective on the organizational and strategic aspects of projects, with over 6,398 questions. Lastly, GenAI (launched in March 2023) is a rapidly growing platform dedicated to the latest developments in AI, particularly in generative models, with 268 questions.

\begin{table}[h!]
\centering
\begin{tabular}{|p{1.5cm}|p{1.2cm}|p{1.2cm}|p{1.2cm}|p{1.2cm}|}
\hline
\textbf{Platforms} & \textbf{Total Questions} & \textbf{Total Answers} & \textbf{Number of Users} & \textbf{Launch Date} \\ \hline
\textbf{Data Science}           & 36,775                   & 55,800                  & 15,900                   & July 2014            \\ \hline
\textbf{Artificial Intelligence} & 12,380                   & 18,200                  & 7,500                    & June 2016            \\ \hline
\textbf{Project Management}     & 6,398                    & 10,100                  & 4,300                    & December 2010        \\ \hline
\textbf{Generative AI (GenAI)}  & 268                      & 420                     & 200                      & March 2023           \\ \hline
\textbf{Software Engineering}   & 63,423                   & 105,000                 & 28,300                   & August 2010          \\ \hline
\end{tabular}
\caption{Overview of Selected Stack Exchange Platforms: Questions, Answers, Users, and Launch Dates}
\label{tab:stackexchange_overview}
\end{table}
\subsection{Questioner Responder (QR) Network}
We have constructed Questioner Responder (QR) Network to observe the user interactions within Stack Exchange. Our approach involves defining the network graph as G =(V,E, W), where V signifies nodes representing users,  E denotes edges symbolizing interactions that link these nodes and W denotes weight. Each user in the network is represented as a unique node, identified by their distinct User ID from the Stack Exchange platform. The User ID acts as a unique identifier for each node, ensuring that all interactions (edges) between users can be accurately tracked and mapped within the graph.  The interactions among users, such as asking or answering questions, are defined by directed edges within the graph. This direct edge represents the flow of knowledge, with an questioner's node connected to the corresponding answerer's node. Moreover, we have integrated the concept of weighted edges into our network graph to encapsulate the significance of user connections. The assignment of weights to edges is influenced by response timeliness. Rapid responses are indicative of a more immediate and engaged interaction, thereby potentially signifying a stronger connection between users. By incorporating these weighted edges into our network representation, we attain a more nuanced and profound understanding of network dynamics. The varying intensities of connections, as indicated by the weights, shed light on the quality of user engagements within the Stack Exchange community, enriching our analysis and interpretation of user interaction patterns. Here, the response time (r) is defined as the temporal difference between the moment a question is posed by the questioner and the subsequent response provided by the responder. This metric is crucial in understanding the efficiency and effectiveness of information dissemination and knowledge exchange dynamics within the network. \\
The edge weight computation formula utilized in our network construction is expressed as follows:
\[
\text{edge weight} = \frac{1}{r+\epsilon}
\]
In this formula, r represents the response time between when a question is posed and when it is answered. We chose to use the reciprocal of the response time,  to ensure that faster responses, which indicate more immediate interactions, result in higher edge weights. This approach allows us to assign greater importance to quicker exchanges, which are often more reflective of active participation and strong connections between users. \\
The addition of $\epsilon = 0.01$ in the denominator serves a practical purpose: it prevents division by zero when the response time is extremely short (or theoretically zero). Without this small adjustment, the formula could produce undefined or disproportionately large weights for very rapid responses. \\
\begin{table}[h!]
\centering
\begin{tabular}{|p{2.2cm}|p{1.5cm}|p{1.5cm}|}
\hline
\textbf{StackExchange Platform}  & \textbf{Number of Nodes} & \textbf{Number of Edges} \\ \hline
\textbf{Data Science}                         & 17,523                   & 26,509                   \\ \hline
\textbf{Artificial Intelligence}                   & 5295                    & 7,546                    \\ \hline
\textbf{Project Management}                         & 4,097                    & 5,700                    \\ \hline
\textbf{Generative AI (GenAI)}                      & 155                      & 133                      \\ \hline
\textbf{Software Engineering}                     & 34,013                   & 57,391                   \\ \hline
\end{tabular}
\caption{Summary of Stack Exchange Platforms: Total Questions, Number of Nodes, and Number of Edges}
\label{tab:stackexchange_summary}
\end{table}

\section{Network Analysis}
The table \ref{tab:stackexchange_summary} provides a summary of five Stack Exchange platforms, outlining the total number of questions, nodes, and edges in each network. The number of nodes represents the unique users involved, while the edges indicate interactions (e.g., questions and answers exchanged) between these users. A key observation from the data is the relatively sparse network structure of GenAI, which has a low number of nodes (155) and edges (133) compared to other platforms. This indicates that user engagement and interactions in GenAI are still developing, likely due to its relatively recent launch. On the other hand, Data Science and Software Engineering exhibit densely connected networks with a significantly higher number of nodes and edges. Data Science, with 17,523 nodes and 26,509 edges, shows a strong level of user interaction and community engagement. Software Engineering, the largest platform in the dataset, has 34,013 nodes and 57,391 edges, reflecting a well-established and highly interactive community. Overall, the table highlights the varying levels of engagement and connectivity across different Stack Exchange platforms, from newer and more niche platforms like GenAI to more mature and densely connected ones like Data Science and Software Engineering.

The table \ref{tab:stackexchange_roles_with_ratio} provides a detailed breakdown of user roles across five Stack Exchange platforms. The roles are divided into three categories: users who only ask questions ("Questioners Only"), users who only provide answers ("Responders Only"), and users who engage in both asking and answering questions ("Both Questioners and Responders").
The GenAI platform, with its relatively small user base, consists of 91 users who are questioners only, 42 users who act solely as responders, and 21 users who participate in both roles. This contrasts sharply with the Software Engineering community, which is the largest among the platforms analyzed. Software Engineering hosts 27,345 users who are exclusively questioners, 6,668 users who only respond to questions, and 3,654 users who are active in both asking and answering questions. This indicates a highly active and engaged user base within the Software Engineering community.
The AI community features 4,128 users who are questioners only, 1,167 responders only, and 685 users who participate in both roles. These figures suggest a balanced level of engagement, with a substantial portion of users contributing to both asking and answering questions. In the Data Science community, there are 14,084 users who are questioners only, 3,448 who are responders only, and 2,074 users who engage in both activities. This indicates a strong level of user engagement, particularly among those who contribute in both capacities.
The Project Management platform, while smaller in comparison to others, shows 3,440 users who are questioners only, 657 who are responders only, and 338 who engage in both roles. The relatively lower numbers in this community may reflect its more specialized focus or smaller user base. Overall, the data illustrate the varying levels of user participation across different platforms, with larger communities like Software Engineering and Data Science showing broader engagement, while more specialized communities like GenAI and Project Management demonstrate focused, yet significant, user interaction. Understanding these dynamics is crucial for tailoring community management strategies to foster greater user interaction and participation.

The \textbf{QR Ratio} column in the table \ref{tab:stackexchange_roles_with_ratio} provides an insight into the balance between questioners and responders across different Stack Exchange platforms. The QR Ratio is calculated by comparing the number of users who ask questions (questioners) to those who provide answers (responders). A higher QR Ratio indicates a larger proportion of questioners relative to responders, which may suggest a higher demand for responses or a lower level of active participation in answering questions. For instance, Project Management exhibits the highest QR Ratio of 5.23, indicating a significant number of users primarily asking questions compared to those answering them. In contrast, GenAI has the lowest QR Ratio of 2.17, suggesting a more balanced interaction between questioners and responders. The Data Science and Software Engineering platforms show similar QR Ratios (4.09 and 4.10, respectively), reflecting a substantial engagement from questioners but also highlighting a need for more active responders. The AI platform, with a QR Ratio of 3.54, also indicates a predominance of questioners, though the gap is slightly narrower compared to other platforms. Overall, the QR Ratios reveal differing dynamics in user engagement, with some platforms experiencing a greater imbalance between questioners and responders than others.

\begin{figure}[h]
  \centering
  \includegraphics[width=0.5\textwidth, height=0.4\textheight]{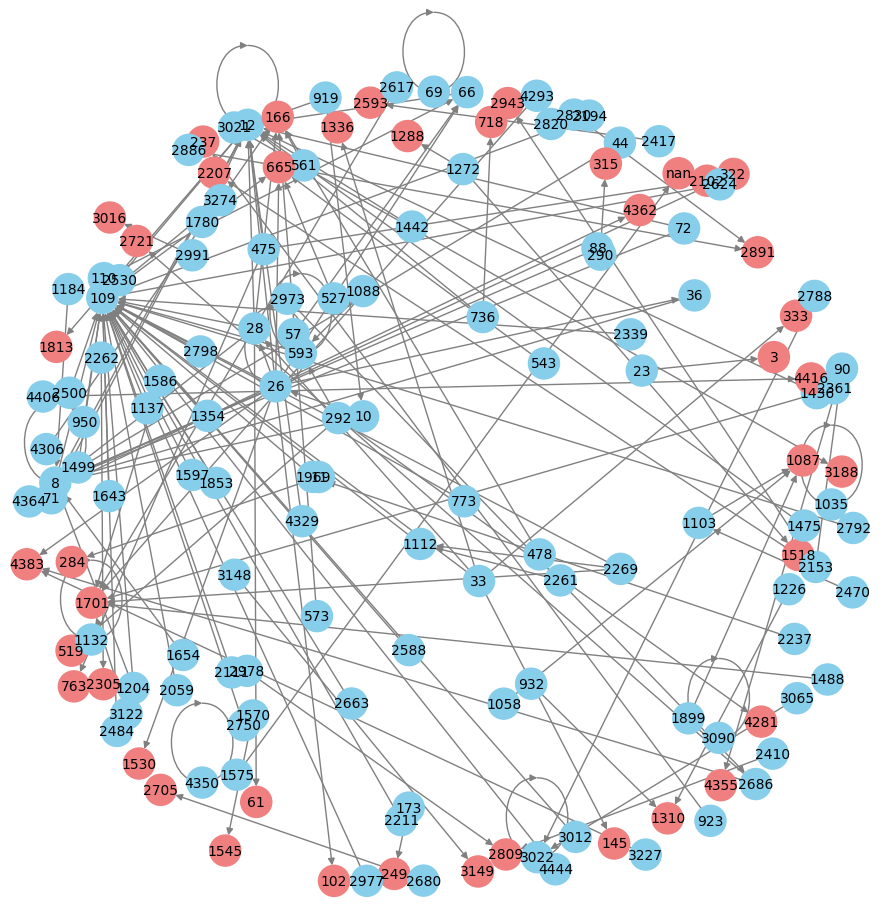}
  \caption{Visualization of the Questioner-Responder Network from the GenAI StackExchange Platform. Nodes represent users, with directed edges indicating the flow of questions and responses. Source nodes (in blue) are users who asked questions, while target nodes (in coral) are users who provided responses. This graph illustrates the interaction dynamics within the community}
  \label{fig:example_image}
\end{figure}

\begin{table}[h!]
\centering
\begin{tabular}{|p{2.2cm}|p{1.2cm}|p{1.2cm}|p{1.2cm}|p{1.2cm}|}
\hline
\textbf{StackExchange Platform} & \textbf{Number of Questioners Only} & \textbf{Number of Responders Only} & \textbf{Number of Both Questioners and Responders} & \textbf{QR Ratio} \\ \hline
\textbf{Data Science}           & 14,084                          & 3,448                          & 2,074                                           & 4.09 \\ \hline
\textbf{Artificial Intelligence} & 4,128                           & 1,167                          & 685                                             & 3.54 \\ \hline
\textbf{Project Management}     & 3,440                           & 657                            & 338                                             & 5.23 \\ \hline
\textbf{GenAI}                  & 91                              & 42                             & 21                                              & 2.17 \\ \hline
\textbf{Software Engineering}   & 27,345                          & 6,668                          & 3,654                                           & 4.10 \\ \hline

\end{tabular}
\caption{Summary of Questioners, Responders, Users Engaged in Both Activities, and the Questioners/Responders Ratio Across Stack Exchange Platforms}
\label{tab:stackexchange_roles_with_ratio}
\end{table}

\section{Statistical Analysis of Centrality Measures Across Stack Exchange Platforms}
In this study, we aim to understand how different users within Stack Exchange platforms contribute to the flow of information and how their roles and influence can be quantified using centrality measures. Centrality metrics such as Degree Centrality, Betweenness Centrality, Closeness Centrality, PageRank, Eigenvector Centrality, and Harmonic Centrality are used to measure various aspects of user engagement and connectivity within the network. By calculating the mean and standard deviation of these metrics across five platforms—GenAI, Software Engineering, AI, Data Science, and Project Management—we can gain insights into how influence and user connectivity vary across different communities. This analysis allows us to identify patterns of interaction, recognize key contributors, and understand the underlying structure of each platform's network. These findings are crucial for interpreting the results, as they provide a quantitative foundation for understanding user dynamics and the effectiveness of knowledge dissemination within these online communities.
\textbf{Degree Centrality} measures the number of direct connections a user has, reflecting their immediate reach or activity level within the network; users with high degree centrality are typically well-connected and involved in numerous interactions. \textbf{Betweenness Centrality} assesses how often a user acts as a bridge between other users, indicating their role in facilitating communication and information flow across different parts of the network; high betweenness centrality users are key connectors who can influence the spread of information. \textbf{Closeness Centrality} calculates how quickly a user can interact with all other users in the network, measuring their efficiency in spreading information; users with high closeness centrality are central to the network and can disseminate or gather information rapidly. \textbf{PageRank} evaluates a user’s influence by considering not just the number of their connections, but also the importance of those they are connected to, identifying users who are linked to other influential users, which amplifies their importance. \textbf{Eigenvector Centrality} extends this idea by considering a user’s connections and the centrality of those connections, highlighting users who hold influence through their association with other highly central users. Finally, \textbf{Harmonic Centrality} offers a variation of closeness centrality that is more robust in disconnected networks, measuring the ease with which a user can be reached by others, making it particularly useful for identifying accessible or influential users across various sub-communities. These centrality measures collectively provide a comprehensive view of user positioning within the network, revealing their potential to influence, connect, and engage with others in meaningful ways, which is crucial for understanding the dynamics of user engagement across different Stack Exchange platforms.
\begin{table*}[h!]
\centering
\begin{tabular}{|p{2cm}|p{2cm}|p{2cm}|p{2cm}|p{2cm}|p{2cm}|p{2cm}|}
\hline
\textbf{Platform} & \textbf{Degree Centrality (Mean ± Std)} & \textbf{Betweenness Centrality (Mean ± Std)} & \textbf{Closeness Centrality (Mean ± Std)} & \textbf{PageRank (Mean ± Std)} & \textbf{Eigenvector Centrality (Mean ± Std)} & \textbf{Harmonic Centrality (Mean ± Std)} \\ \hline
\textbf{Data Science} & 0.0001 ± 0.0007 & 0.0000 ± 0.0001 & 0.0014 ± 0.0081 & 0.0001 ± 0.0004 & 0.0005 ± 0.0071 & 30.4983 ± 177.7957 \\ \hline
\textbf{Artificial Intelligence} & 0.0004 ± 0.0018 & 0.0000 ± 0.0001 & 0.0017 ± 0.0099 & 0.0002 ± 0.0026 & 0.0010 ± 0.0128 & 11.6321 ± 69.4624 \\ \hline
\textbf{Project Management} & 0.0006 ± 0.0026 & 0.0000 ± 0.0002 & 0.0048 ± 0.0235 & 0.0002 ± 0.0016 & 0.0017 ± 0.0149 & 23.4470 ± 115.8650 \\ \hline
\textbf{GenAI} & 0.013 ± 0.0236 & 0.0001 ± 0.0006 & 0.009 ± 0.0308 & 0.0065 ± 0.0358 & 0.0065 ± 0.0808 & 1.5142 ± 5.2904 \\ \hline
\textbf{Software Engineering} & 0.0001 ± 0.0005 & 0.0000 ± 0.0001 & 0.0032 ± 0.0144 & 0.0000 ± 0.0004 & 0.0004 ± 0.0051 & 144.5411 ± 639.3517 \\ \hline
\end{tabular}
\caption{Centrality Measures (Mean ± Standard Deviation) Across Stack Exchange Platforms. The table presents the average values and variability for Degree Centrality, Betweenness Centrality, Closeness Centrality, PageRank, Eigenvector Centrality, and Harmonic Centrality, highlighting the differences in user influence and connectivity across different platforms.}
\label{tab:centrality_combined}
\end{table*}



\subsection{Description of Centrality Correlation Matrices}

The correlation matrices for centrality scores (as shown in table \ref{tab:centrality_correlation}) across five different Stack Exchange platforms—Data Science, AI, Project Management, GenAI, and Software Engineering—provide a detailed understanding of the relationships between various centrality measures within each network. These centrality measures include Degree Centrality, Betweenness Centrality, PageRank, Closeness Centrality, Harmonic Centrality, and Eigenvector Centrality. The correlations between these measures reveal insights into the structure and dynamics of user interactions within each community.

\subsection{Data Science}
In the Data Science Stack Exchange community, the correlations between most centrality measures are relatively moderate. The strongest correlations are observed between Degree Centrality and PageRank (0.6957) and between Degree Centrality and Eigenvector Centrality (0.6437). This indicates that users who are well-connected (high Degree Centrality) also tend to be influential (high PageRank) and have connections to other influential users (high Eigenvector Centrality). However, the correlation between Betweenness Centrality and other measures is generally weaker, suggesting that the role of users as intermediaries or bridges in the network is less aligned with their direct connections or influence.

\subsection{Artificial Intelligence}
The AI Stack Exchange network exhibits a similar pattern to the Data Science community, with strong correlations between Degree Centrality, Betweenness Centrality, and PageRank. For instance, the correlation between Betweenness Centrality and PageRank is particularly high (0.7284), indicating that users who frequently act as bridges in the network also hold significant influence. However, the correlations involving Harmonic Centrality are very low or negative, suggesting that the ease of reaching a user (Harmonic Centrality) does not align with other forms of centrality in this community.

\subsection{Project Management}
The Project Management Stack Exchange network shows some of the highest correlations between centrality measures. Notably, the correlation between Degree Centrality and PageRank is extremely high (0.8751), as is the correlation between Degree Centrality and Betweenness Centrality (0.7795). These strong correlations indicate a tightly interconnected network where users with many direct connections also tend to be influential and central to the flow of information. The relatively strong correlation between Betweenness Centrality and Eigenvector Centrality (0.7221) further suggests that key users often serve both as influential nodes and as bridges between different parts of the network.

\subsection{Generative AI}
In the GenAI community, there is a high correlation between Degree Centrality and Betweenness Centrality (0.8927), indicating that users who are well-connected also play significant intermediary roles. However, the correlation between Degree Centrality and PageRank is lower (0.4994) compared to other communities, suggesting that having many connections does not necessarily translate into overall influence within this community. Interestingly, the correlation between Closeness Centrality and Degree Centrality is relatively high (0.8034), implying that users who are central in terms of direct connections are also efficient in spreading information.

\subsection{Software Engineering}
The Software Engineering Stack Exchange network exhibits a strong alignment between Degree Centrality, Betweenness Centrality, and PageRank, with correlations of 0.6881 and 0.7511 between Degree Centrality and PageRank, and between Betweenness Centrality and PageRank, respectively. These strong correlations suggest that users who are well-connected and act as bridges also hold significant influence. The correlation between Degree Centrality and Eigenvector Centrality is particularly high (0.7079), indicating that users with many connections are not only influential but are also connected to other influential users.

\subsection{Variations Across Communities}
Across all five platforms, Degree Centrality tends to correlate strongly with PageRank and Eigenvector Centrality. This suggests that users who are well-connected are generally influential, both in terms of direct connections and in their association with other influential users. The Project Management community stands out with exceptionally high correlations between most centrality measures, indicating a tightly-knit and hierarchical network structure. In contrast, the Data Science and AI communities show weaker correlations involving Betweenness Centrality, implying that the role of intermediaries in these networks is less central to overall influence. The GenAI community’s relatively lower correlation between Degree Centrality and PageRank highlights a more distributed or egalitarian network, where having many connections does not necessarily equate to high influence.

\subsection{Harmonic Centrality}
Across all platforms, Harmonic Centrality shows weak correlations with other centrality measures, particularly in Data Science and AI. This suggests that the ease of reaching a user (Harmonic Centrality) is not strongly tied to their centrality or influence within these networks. This metric might be less relevant in analyzing user influence in these communities. \\

The correlation analysis of centrality scores across the five Stack Exchange platforms reveals both common patterns and unique characteristics of each community. While certain centrality measures consistently correlate across platforms, reflecting general trends in user influence and connectivity, the variations in specific correlations provide insights into the distinct network structures and dynamics of each community. These findings can inform strategies for enhancing user engagement and optimizing knowledge dissemination within these online communities.
\begin{table*}[]
\resizebox{\textwidth}{!}{%
\begin{tabular}{|lrrrrrr|}
\hline
\multicolumn{7}{|c|}{\textbf{Data Science}} \\ \hline
\textbf{Centrality Measure} & \textbf{Degree Centrality} & \textbf{Betweenness Centrality} & \textbf{PageRank} & \textbf{Closeness Centrality} & \textbf{Harmonic Centrality} & \textbf{Eigenvector Centrality} \\ \hline
Degree Centrality           & 1.000000                   & 0.522537                       & 0.695683         & 0.404590                     & 0.016771                     & 0.643708                       \\ \hline
Betweenness Centrality      & 0.522537                   & 1.000000                       & 0.437860         & 0.304454                     & 0.002910                     & 0.548558                       \\ \hline
PageRank                   & 0.695683                   & 0.437860                       & 1.000000         & 0.384238                     & 0.006777                     & 0.692014                       \\ \hline
Closeness Centrality        & 0.404590                   & 0.304454                       & 0.384238         & 1.000000                     & 0.023643                     & 0.605168                       \\ \hline
Harmonic Centrality         & 0.016771                   & 0.002910                       & 0.006777         & 0.023643                     & 1.000000                     & 0.015892                       \\ \hline
Eigenvector Centrality      & 0.643708                   & 0.548558                       & 0.692014         & 0.605168                     & 0.015892                     & 1.000000                       \\ \hline
\multicolumn{7}{|c|}{\textbf{Artificial Intelligence}} \\ \hline
\textbf{Centrality Measure} & \textbf{Degree Centrality} & \textbf{Betweenness Centrality} & \textbf{PageRank} & \textbf{Closeness Centrality} & \textbf{Harmonic Centrality} & \textbf{Eigenvector Centrality} \\ \hline
Degree Centrality           & 1.000000                   & 0.566556                       & 0.671766         & 0.494718                     & -0.002473                    & 0.657648                       \\ \hline
Betweenness Centrality      & 0.566556                   & 1.000000                       & 0.728442         & 0.400395                     & -0.002661                    & 0.647429                       \\ \hline
PageRank                   & 0.671766                   & 0.728442                       & 1.000000         & 0.292062                     & -0.002817                    & 0.581229                       \\ \hline
Closeness Centrality        & 0.494718                   & 0.400395                       & 0.292062         & 1.000000                     & 0.000629                     & 0.672961                       \\ \hline
Harmonic Centrality         & -0.002473                  & -0.002661                      & -0.002817        & 0.000629                     & 1.000000                     & -0.003336                      \\ \hline
Eigenvector Centrality      & 0.657648                   & 0.647429                       & 0.581229         & 0.672961                     & -0.003336                    & 1.000000                       \\ \hline
\multicolumn{7}{|c|}{\textbf{Project Management}} \\ \hline
\textbf{Centrality Measure} & \textbf{Degree Centrality} & \textbf{Betweenness Centrality} & \textbf{PageRank} & \textbf{Closeness Centrality} & \textbf{Harmonic Centrality} & \textbf{Eigenvector Centrality} \\ \hline
Degree Centrality           & 1.000000                   & 0.779460                       & 0.875104         & 0.454167                     & 0.058998                     & 0.571566                       \\ \hline
Betweenness Centrality      & 0.779460                   & 1.000000                       & 0.746989         & 0.447518                     & 0.069468                     & 0.722136                       \\ \hline
PageRank                   & 0.875104                   & 0.746989                       & 1.000000         & 0.459117                     & 0.042802                     & 0.601911                       \\ \hline
Closeness Centrality        & 0.454167                   & 0.447518                       & 0.459117         & 1.000000                     & 0.102534                     & 0.685860                       \\ \hline
Harmonic Centrality         & 0.058998                   & 0.069468                       & 0.042802         & 0.102534                     & 1.000000                     & 0.122864                       \\ \hline
Eigenvector Centrality      & 0.571566                   & 0.722136                       & 0.601911         & 0.685860                     & 0.122864                     & 1.000000                       \\ \hline
\multicolumn{7}{|c|}{\textbf{Software Engineering}} \\ \hline
\textbf{Centrality Measure} & \textbf{Degree Centrality} & \textbf{Betweenness Centrality} & \textbf{PageRank} & \textbf{Closeness Centrality} & \textbf{Harmonic Centrality} & \textbf{Eigenvector Centrality} \\ \hline
Degree Centrality           & 1.000000                   & 0.640932                       & 0.688059         & 0.313039                     & 0.012672                     & 0.707919                       \\ \hline
Betweenness Centrality      & 0.640932                   & 1.000000                       & 0.751057         & 0.267754                     & 0.015129                     & 0.660915                       \\ \hline
PageRank                   & 0.688059                   & 0.751057                       & 1.000000         & 0.177912                     & 0.004359                     & 0.620245                       \\ \hline
Closeness Centrality        & 0.313039                   & 0.267754                       & 0.177912         & 1.000000                     & 0.048700                     & 0.487333                       \\ \hline
Harmonic Centrality         & 0.012672                   & 0.015129                       & 0.004359         & 0.048700                     & 1.000000                     & 0.018744                       \\ \hline
Eigenvector Centrality      & 0.707919                   & 0.660915                       & 0.620245         & 0.487333                     & 0.018744                     & 1.000000                       \\ \hline
\multicolumn{7}{|c|}{\textbf{Generative AI}} \\ \hline
\textbf{Centrality Measure} & \textbf{Degree Centrality} & \textbf{Betweenness Centrality} & \textbf{PageRank} & \textbf{Closeness Centrality} & \textbf{Harmonic Centrality} & \textbf{Eigenvector Centrality} \\ \hline
Degree Centrality           & 1.000000                   & 0.892730                       & 0.499418         & 0.803361                     & -0.027601                    & 0.291489                       \\ \hline
Betweenness Centrality      & 0.892730                   & 1.000000                       & 0.632889         & 0.740877                     & -0.034627                    & -0.011850                      \\ \hline
PageRank                   & 0.499418                   & 0.632889                       & 1.000000         & 0.606185                     & -0.027619                    & 0.159168                       \\ \hline
Closeness Centrality        & 0.803361                   & 0.740877                       & 0.606185         & 1.000000                     & 0.000046                     & 0.393652                       \\ \hline
Harmonic Centrality         & -0.027601                  & -0.034627                      & -0.027619        & 0.000046                     & 1.000000                     & -0.007788                      \\ \hline
Eigenvector Centrality      & 0.291489                   & -0.011850                      & 0.159168         & 0.393652                     & -0.007788                    & 1.000000                       \\ \hline
\end{tabular}%
}
\caption{Correlation Matrix of Centrality Measures Across Stack Exchange Platforms
This table presents the correlation coefficients between various centrality measures (Degree Centrality, Betweenness Centrality, PageRank, Closeness Centrality, Harmonic Centrality, Eigenvector Centrality) across different Stack Exchange platforms, including Data Science, AI, Project Management, Software Engineering, and GenAI. The correlations highlight the relationships between user influence, connectivity, and interaction strength within each community.}
\label{tab:centrality_correlation}
\end{table*}

\section{Results}

The detailed analyses of user engagement patterns within various Stack Exchange communities reveal significant insights on how influence and contributions vary among users. By constructing a Questioner-Responder (QR) network, users were represented as nodes, and directed edges captured interactions between users, with the edge weight being a function of the response time. This study highlights the importance of edge weights determined by response time; faster interactions often signify stronger connections within the network. Centrality measures (i.e., Degree Centrality, Betweenness Centrality, Closeness Centrality, PageRank, Eigenvector Centrality, Harmonic Centrality) were crucial in identifying key contributors across different platforms.

The comparative analysis across platforms, including Data Science, AI, Project Management, GenAI, and Software Engineering, revealed diverse engagement dynamics. For instance, smaller communities including Gen AI and Project Management exhibited high variability in centrality measures, indicating a concentration of influence among a few users. In contrast, larger communities like Software Engineering demonstrated a more distributed network structure with low centrality scores and many edges. The size of any user-base has a significant impact on user role distribution (questioners, responders, and both-ways). Software Engineering showed a different engagement dynamics, than GenAI. On the other hand, Project Management and GenAI displayed strong correlations between centrality measures and interaction strength, suggesting tightly-knit networks. In contrast, platforms like Data Science and AI showed weaker correlations, implying a more dispersed engagement pattern. These findings underscore the need for tailored strategies to enhance user interaction and knowledge dissemination based on the unique characteristics of each community.

\subsection{RQ1: User Roles and Engagement Patterns}

User roles and engagement patterns vary across different stack exchange platforms. For instance, the Software engineering community with a large user base displayed a higher proportion of both questioners and responders compared to smaller communities such as Gen AI (Table 1). Therefore, larger platforms tend to foster more balanced participation while smaller or specialized platforms may focus on specific roles (i.e., asking or answering questions). Engagement was also strongly correlated with user centrality measures, such as Degree centrality and Betweenness Centrality. This implied that users with high centrality scores play a pivotal role in maintaining the flow of information.



\subsection{RQ2: Influential Users}

By examining various centrality measures across the platforms, we identified key users who hold significant influence in their respective communities. These key users typically have higher Degree Centrality which denotes many direct connections and high PageRank as of their influence within the network. In specialized communities such as Gen AI and Project Management, a small number of users exhibited both high Degree and Betweenness Centrality, suggesting that these users act as bridges between different parts of the network. Conversely, in more populated communities such as Software Engineering, key user influence is more distributed with low average centrality scores. The statistical analysis of centrality measures across platforms further revealed that communities with a higher proportion of influential users tend to have stronger correlations between centrality measures and interaction strength. For instance, in Project Management, users with high centrality scores were more likely to engage in stronger and immediate interactions, reinforcing the cohesive nature of this community (Table 2).

\subsection{RQ3: Comparative Community Dynamics}

The dynamics within different Stack Exchange communities vary greatly based on their focus and user base. For example, the Data Science community displayed a lower correlation between centrality measures and edge weights, suggesting that the strength of user connections is less influenced by centrality in this field. In contrast, Gen AI community, despite its smaller size exhibited strong positive correlations between user centrality and engagement metrics, indicating a more interconnected and active user base. These findings suggest that different communities may require tailored strategies for fostering engagement. The exploration of cross-platform interactions, where users engage in multiple platforms simultaneously, could offer new insights into user behavior dynamics. 

\section{Limitations and Future Research}
The current study has several limitations. First, the proposed QR Network approach offers insights into user interactions but it does not account for nuanced engagement patterns such as passive interactions (e.g., users who only view content without contributing). Additionally, the study does not factor in external influences on engagement, such as evolving community policies or the impact of emerging technologies (e.g., AI tools like ChatGPT) on user behavior. Future research could capture passive engagements to any platform's ecosystem and shed light on this segment through qualitative analyses. Morever, future work could investigate the longitudinal changes in user engagement to understand how evolving community norms or external technological advancements (e.g., AI-driven tools) influence participation over time. Second, the current scope is limited to a few Stack Exchange communities (e.g., Data Science, AI, Software Engineering). Expanding to other Stack Exchange communities or similar platforms would provide a holistic understanding of user engagement patterns. .  Third, our analyses primarily focus on centrality measures to identify key influencers, but this method might overlook other dimensions of influence, such as domain expertise or the quality of contributions.  Besides, the reliance on computational metrics (e.g., response time, edge weights) to assess engagement may not fully capture the qualitative aspects of user behavior, such as the content's complexity or users' intent behind their interactions. Future research could explore the role of content quality and domain expertise in driving user influence, moving beyond purely quantitative metrics like centrality scores.  Fourth, the dataset utilized for network analysis is limited by the archival nature of Stack Exchange data. Although these datasets are comprehensive and anonymized but may not accurately represent real-time engagement trends. Incorporating predictive models to forecast engagement trends could be a valuable extension, particularly for community managers seeking to sustain or revitalize user participation. 

\section{Conclusion}
The network constructed from questioner-responder interactions provides valuable insights into user engagement, laying the foundation for enhancing user experience and fostering a more knowledgeable community. By applying network metrics such as centrality (user-based), and edge weight, we can better understand user behavior and interaction patterns. The analysis of the combined graph uncovers important findings regarding user relationships, and community structures within the dataset.

The observed dense connectivity and moderate degree centrality suggest a cohesive network of users with shared interests. This creates opportunities for improving community engagement, offering personalized content, and informing strategic decisions based on network dynamics.

This framework can be applied to several key areas. Identifying users at risk of disengagement or churn becomes possible by observing low centrality scores or users leaving tightly-knit communities. In addition, users with high centrality, particularly in terms of betweenness centrality, can be recommended as experts to answer questions in relevant domains. Furthermore, identifying users with complementary skill sets through non-overlapping connections can promote collaboration on complex or interdisciplinary questions.

By leveraging these insights, it is possible to drive more effective engagement, increase expert participation, and foster a collaborative environment that enhances the overall value of the community.

 \bibliographystyle{IEEEtran}
 \bibliography{mybibliography}


\end{document}